# Incoherent Optoelectronic Differentiation with Optimized Multilayer Films

Xiaomeng Zhang[1,2], Benfeng Bai[*, 1], Hong-bo Sun[1], Guofan Jin[1], Jason Valentine[*, 2]

[1] State Key Laboratory of Precision Measurement Technology and Instruments, Department of Precision Instrument, Tsinghua University, Beijing 100084, China
[2] Department of Mechanical Engineering, Vanderbilt University, Nashville, TN, USA.
Email: baibenfeng@tsinghua.edu.cn, jason.g.valentine@vanderbilt.edu

**Abstract**

Fourier-based optical computing operations, such as spatial differentiation, have recently been realized in compact form factors using flat optics. Experimental demonstrations, however, have been limited to coherent light requiring laser illumination and leading to speckle noise and unwanted interference fringes. Here, we demonstrate the use of optimized multilayer films, combined with dual color image subtraction, to realize differentiation with unpolarized incoherent light. Global optimization is achieved by employing neural networks combined with the reconciled level set method to optimize the optical transfer functions of multilayer films at wavelengths of 532 nm and 633 nm. Spatial differentiation is then achieved by subtracting the normalized incoherent images at these two wavelengths. The optimized multilayer films are experimentally demonstrated to achieve incoherent differentiation with a numerical aperture up to 0.8 and a resolution of 6.2 μm. The use of multilayer films allows for lithography-free fabrication and is easily combined with existing imaging systems opening the door to applications in microscopy, machine vision and other image processing applications.

With the rapidly growing demands for high performance computing of large datasets, such as tensor computing in image processing and deep learning, researchers have returned to optical analog computing as a complementary technique to digital processing [1] due to its inherent parallelism, low energy consumption and high processing speed. As an example, recent demonstrations of hybrid optical and digital approaches such as reconfigurable neuromorphic computing [2] and vector-matrix multiplication [3] have been experimentally demonstrated to provide a ~$10^5$ times increase in energy efficiency and ~5 times increase in computing speed. However, optical analog computing operations [4] based on conventional optical elements, such as holograms and spatial light modulators (SLMs), result in bulky systems that are difficult to align and integrate into compact systems. Recently, metamaterials and metasurfaces have been utilized to significantly shrink the size of Fourier-based optical computing systems [5–7] in moving towards integrated image processing systems. Nonlocal metasurfaces [8,9] with functions engineered in the wave vector domain, can be used to further reduce the size of the system by avoiding the need for a Fourier transform. In this case, a metasurface can achieve certain Fourier-based optical computing functions, such as edge detection, by being directly inserted into a standard coherent imaging system.

Edge detection is often the first step in image processing, as it allows for segmentation of objects. The technique also has applications in microscopy such as cellular imaging, as it can be used for cell boundary identification, phenotypic monitoring, and disease diagnosis [10]. By exploiting deep learning in microscopy, tasks such as cell counting, segmentation, tracking and identification no longer need task-oriented codes but can be achieved by generalized and versatile neural networks that are both fast and robust [11,12]. The learning features of the first layers of these conventional image-processing neural networks, whose inputs are usually bright-field cellular images, typically represent the presence or absence of edges at particular orientations and locations in the images [13,14]. The use of optical edge images, instead of bright-field images, can save time and energy by replacing the neural network layers that normally perform these functions.

To date, experimental demonstrations of metasurface-based optical edge detectors require spatially coherent light and thus suffer from laser speckle and interference fringes resulting in sensitivity to imperfections in the optical path. Incoherent differentiation can avoid these problems and has been traditionally achieved by building bipolar point spread functions (PSFs) followed by a digital subtraction. Bipolar PSFs have traditionally been achieved by employing two filters at the Fourier plane of a $4f$ correlator or by using two separate holographic apertures. These approaches require spatial or temporal multiplexing of separate apertures, complicating the optical systems and resulting in image mismatch originating from separate apertures [15]. Recently, a compact incoherent optoelectronic differentiation scheme based on frequency multiplexing with photonic crystals [16] was proposed. This approach requires one aperture and significantly shrinks the size of the system but would involve multilayer photonics crystals and is limited to one-dimensional (1D) differentiation with polarized light and closely spaced working wavelengths.

Here, we propose and experimentally demonstrate a single aperture, incoherent optoelectronic differentiation scheme, based on optimized multilayer films that can be directly inserted into conventional imaging systems, as shown in Fig. 1. The optical transfer functions (OTFs) of the multilayer films at the wavelengths of 633 nm and 532 nm, and for both TE and TM polarization, are optimized to ensure that the difference in the OTFs at these two wavelengths results in second order differentiation. In this scheme, edge detection is achieved by subtracting the normalized incoherent images at these two wavelengths. We optimize the OTFs of the multilayer films with a global optimization neural network (GLOnet) [17] together with the reconciled level set method [18]. The deliberately designed polarization insensitive multilayer films convert the convolution operation required by electric incoherent differentiation to a subtraction of two normalized incoherent images at these two wavelengths, so that the computational consumption is lowered by a factor of 10 [16]. Furthermore, the incoherent images can be distinguished by the color filters already placed on conventional detectors working in visible regime. Compared to other nanophotonic structures such as metasurfaces, the multilayer films can provide an azimuthally uniform response for 2D edge detection and can be readily fabricated over large areas without the need for lithographic patterning. This compact scheme of polarization-insensitive incoherent optoelectronic differentiation can significantly broaden the capacity of optical image processing.

Edge detection with incoherent light is distinct from coherent light [19] as with the former, the fields at different locations are uncorrelated and thus the optical system is linear for the intensity of light rather than the amplitude. In the case of incoherent light, the output intensity $I_{out}(k)$ of light at a spatial frequency $k$ is expressed as

$$I_{out}(k) = I_{in}(k)T(k), \qquad (1)$$

where $I_{in}(k)$ is the input intensity and $T(k)$ is the OTF. The relationship between $T(k)$ and the amplitude transfer function (ATF) $t(k)$ is given by

$$T(k) = \frac{1}{N}\int_{-\infty}^{\infty} t^*(k)t(k+q)dq, \qquad (2)$$

where $N$ is an intensity normalization factor. To realize second order differentiation on an input intensity image, we need

$$I_{out}(k) = \int I_{out}(r)e^{ik\cdot r}dr = \int [\nabla^2 I_{in}(r)]e^{ik\cdot r}dr = -k^2 I_{in}(k). \qquad (3)$$

Thus, $T(k)$ is required to be

$$T(k) \propto k^2. \qquad (4)$$

However, the autocorrelation function of $t(k)$, $T(k)$ comes to a maximum at $k=0$, according to the Cauchy-Schwarz inequality, rather than 0 required in Eq. (4). To reconcile this issue, and realize an effective OTF proportional to $k^2$, two different OTFs can be designed at either two different polarizations or two different wavelengths. In this case, digital subtraction of these two OTFs can satisfy the requirement of Eq. (4).

Traditional incoherent differentiation systems often utilize different PSF filters in the Fourier plane for small numerical aperture (NA) imaging systems or separate holographic apertures for larger NAs [20–22] to acquire different OTFs. Modulation between different OTFs is realized by spatial or temporal multiplexing of separate apertures, which complicates the system and may causes image mismatch from apertures that are physically separated. In contrast to these traditional approaches, we propose to use a single aperture comprising a compact multilayer film stack, whose

OTFs under both TE and TM polarizations at the wavelengths of 633 nm and 532 nm are deliberately optimized to ensure that the subtraction of their OTFs is proportional to $k^2$ resulting in a 2D, polarization insensitive, second order differentiation. The two wavelengths are chosen to enable separation of the images using a conventional detector with red, green and blue (RGB) filters, working in the visible spectral range.

Optimization of multilayer films to achieve designer transmittance or reflectance properties has been extensively explored with particle swarm optimization [23], needle optimization [24,25], and memetic algorithms [26] being the most popular. These methods, however, do not consider any local gradient information, resulting in a considerable computational cost to converge. Recently, it has been shown that global optimization can be achieved, at a low computational cost, by combining neural networks and the local gradient provided by the transfer matrix method (TMM) [27,28]. However, the existing optimization techniques based on deep learning have limitations on the configurations of the multilayer films set by the method itself, like the sequence of the materials or the number of layers, which shrinks the design space and impedes finding the optimal structures among all practically feasible ones. Here, we explore the optimization of multilayer films with the B-spline based reconciled level set method embedded in the neural network. This approach avoids limitations caused by the optimization method and achieves a global optimization that can still employ user-defined geometric limitations associated with fabrication constraints.

In order to enable fabrication of the multilayer films, certain restrictions were placed on optimization. First, we consider three materials that are commonly used in photonic devices, $TiO_2$, $HfO_2$ and $SiO_2$, and the number of layers is restricted to be less than 30. The thickness of each layer is also required to be between 30 nm and 300 nm, with a thickness discretization of 5nm. We allow the multilayer films to be of any material sequence as long as the aforementioned restrictions are met. The optimization objective function is to achieve an effective OTF proportional to $k^2$ and a schematic of our optimization process is outlined in Fig.2. The basic structure follows the concept of GLOnets [17]. GlOnets incorporate an electromagnetic simulator (here it is TMM) with a generative neural network to perform the nonconvex global optimization of photonic devices. It does not need any training set of known devices but can learn directly through electromagnetic simulations of the devices that it randomly generates based on the performance and the local gradients of previous design iterations, resulting in a global optimization with modest computational time. The inputs of our conditional generative neural network are the ideal effective OTF and a uniformly distributed noise vector. The intermediate outputs are the B-spline coefficients of different materials in the space domain of 0~9 μm and a percentage thickness factor $h$. The function of $h$ is used to dynamically change the upper limit of the B-spline domain, so that multilayer film stacks with both a thickness less than 9 μm and a number of layers less than 30 are achieved. The final outputs are the B-spline coefficients of different materials truncated by the thickness factor. These outputs allow us to generate multilayer films with a high level of design freedom. Based on the truncated B-spline coefficients and the B-spline basis functions [29], we construct different level set functions for each material. The multilayer films stack is then established according to the reconciled level set method, by choosing the material with the largest value across the domains of the level set function to occupy the corresponding physical positions. The final multilayer film stack undergoes another truncation if its number of layers exceeds 30.

This flexible representation allows the spontaneous changing of materials, thicknesses and number of layers, ranging the entire configuration space of the multilayer films, within the set bounds. In addition, the B-spline basis functions provide not only a linearly superimposed way to construct the level set functions, which makes the truncations possible, but also a preliminary thickness filter for each layer by carefully selecting the degree and the uniform knots number of the B-spline basis functions (see details in Supplementary Note 2).

TMM is used to calculate the ATF of the established thin films under both polarizations at wavelengths of 633 nm and 532 nm. TMM is a fully analytic and high-speed solver that is widely used in multilayer films modeling. Based on these results, the effective OTF is acquired. The thickness limitations of each layer should also be considered and here, instead of applying a geometric penalty function, we use the method of Lagrange multiplier [30] to avoid negative effects on the training process. The Lagrange multiplier together with the rough thickness filtering from the B-spline basis functions ensures that films satisfy the imposed constraints.

The loss function of the neural network with a batch size $A$ is written as

$$L = \sum_{a=1}^{A} -\frac{1}{A} \{ \exp[-(\frac{\sum_k T_{eff}^{(a)}(k) - k^2}{\sigma})^2] + v^{(a)} G^{(a)} \}, \tag{5}$$

where $\sigma$ is a hyperparameter, $v^{(a)}$ is the Lagrange multiplier for the $a$-th multilayer films in the batch, and $G^{(a)}$ is its geometry constraint function that equals to 0 when the thickness of each layer meets the requirements. The Lagrange multiplier $v$ needs to be carefully chosen in each iteration according to the performance of each multilayer film stack in the batch. The details of the loss function are discussed in Supplementary Note 6. The gradient of the loss function with regard to the thickness of each layer is then used to calculate the gradient with regard to the truncated B-spline coefficients, and later the gradient with regard to the neuron weights in the neural network. The neuron weights are then changed accordingly to finish one iteration of the optimization, as shown in Fig. 2. When the training is completed, a batch of multilayer films with much smaller loss values are generated. The optimization method here is designed to be capable of considering fabrication limitations and is flexible enough to adapt to different problems concerning the optimization of multilayer films.

When we apply the optimization method to the design of multilayer films for incoherent optoelectronic differentiation, the films are required to perform identically for both TE and TM polarizations. The NA of the ATF is set to be 0.4, which are common values in microscopy. We expect the effective OTF to approach $k^2$ within a NA of 0.4. With this approach, the optimized structure is a 30-layer aperiodic films stack with a total thickness of 5.5μm, whose optical properties are shown in Fig. 3. It can be seen in Figs. 3(a) and (b) that the ATFs are nearly identical for both TE and TM polarizations at 532 nm and 633 nm, respectively. The OTFs at these two wavelengths are shown in Fig. 3(c) and their subtraction yields the effective OTF shown in Fig. 3(d). However, the multilayer films, limited by the aforementioned fabrication constraints, provide an effective OTF proportional to $k^2$ within a NA of 0.3, as shown in Fig. 3(d).

In order to validate the modeling, the multilayer film stack was deposited using ion assisted E-beam evaporation. A 4*f* correlator was used to image the transmittance of the multilayer film stack under unpolarized light in Fourier space (see polarization-insensitive experimental results in Supplementary Note 3). Comparisons of the simulated and experimental ATFs are provided in Figs. 4(a) and (b). We find that, due to fabrication tolerances, the working wavelengths change from 532 nm and 633 nm to 527nm and 629 nm, respectively. The OTF is measured at these two wavelengths with a rotating diffuser inserted into the optical path to convert coherent light from a super continuum laser into incoherent light [31]. As can be seen from Fig. 4(c) the experimentally measured effective OTF shows agreement with the ideal quadratic function up to an NA of 0.7. This is considerably higher than what is predicted by the modeling presented in Fig. 3 would suggest and is a consequence of coherence remaining in the illumination beam. Note that we assume the light to be completely incoherent in the optimization.

In practice, the degree of coherence in our system is a function of the illumination beam [32], originating from a coherent super-continuum laser, and the diffuser properties. When the optical system is not completely incoherent, but can still work as a linear system with no obvious interference properties, we can approximate the OTF with the following equation [16],

$$T(k) = \frac{1}{N} \int t^*(q) t(k+q) \exp(-\frac{\Delta^2 (k+2q)^2}{4}) dq, \qquad (6)$$

where $\Delta$ is the transverse correlation length which should be small compared with the image size so that the system can still be regarded as a linear one. We fit the measured OTFs according to Eq. (6) and find an approximate transverse correlation length of 4.19 μm at 527 nm and 8.06 μm at 629 nm for our imaging system (see details in Supplementary Note 4). We simulate the effective OTFs with different $\Delta$ to observe the influence of coherence, as shown in Fig. 4(d), where we assume an identical $\Delta$ at both wavelengths. Ultimately, as coherence is improved, there is an increase in $\Delta$ and the effective OTF satisfies second order differentiation within a larger NA. For our imaging system, the measured effective OTF overlaps well with the quadratic reference within a NA of 0.7, as shown in Fig. 4(c). The convolution of an ideal delta function with the effective OTF, along with the Rayleigh criteria, results in a theoretical resolution of 0.6 μm based on the effective OTF [33]. However, the transverse correlation length in our imaging system provides the actual resolution, which is 8.06 μm at a wavelength of 633 nm, a value that is similar to incoherent optoelectronic differentiation with conventional optical elements [20]. The experimental results demonstrate that the optimized films can provide differentiation for a wide variety of incoherent or partially coherent optical systems. Furthermore, if the transversal correlation length of the system is known beforehand, we can directly use this information in the optimization for achieving a more accurate prediction of the effective OTF.

To experimentally demonstrate incoherent optoelectronic spatial differentiation of an object, we imaged both the 1951 USAF resolution test chart and a biological sample with the multilayer film stack inserted into an incoherent imaging system as shown in Fig. 5(a). As before, the incident light is from a super continuum laser filtered to wavelengths of 527 nm and 629 nm. The captured RGB images of Element 3 (Group 4 to Group 6) on the 1951 USAF resolution test chart are shown in Fig. 5(b), together with their corresponding spatial differentiation obtained by digitally subtracting the normalized images in the green and red channels. The transversal correlation length at 629 nm in our system is 8.06μm, which is very close to the image size in Group 6 leading to some blurring in the image. It should be noted that although the resolution of our incoherent optoelectronic differentiation system is experimentally demonstrated as 6.2 μm, the resolution here is not limited by the performance of the multilayer films, but by the transverse correlation length determined by the illumination and collection conditions of the system. We have also imaged a biological sample, i.e., the cross section of a woody dicot stem, as shown in Fig. 5(c). The edge image demonstrates the effectiveness of the multilayer films in 2D, polarization insensitive, incoherent optoelectronic differentiation, which is compact, energy-efficient, and requires no temporal or spatial multiplexing.

In conclusion, we have proposed and experimentally demonstrated a single aperture, polarization insensitive 2D incoherent optoelectronic differentiator where edge detection is achieved by subtraction of normalized incoherent images at the wavelengths of 532 nm and 633 nm. This was achieved by using global optimization of lithography-free multilayer films resulting in an approach that is compact, easy to fabricate and scale, and compatible with conventional imaging systems. Optoelectronic differentiation, in conjunction with optimization methods, provides an avenue to realize incoherent optical computing, significantly decreasing processing time and energy usage, and doing so with compact optical elements. This approach can be particularly useful in microscopy applications such as real-time image processing, object tracking, and disease diagnosis [34,35].

**Acknowledgements**

X.Z. and B.B acknowledges the support by the National Natural Science Foundation of China (NSFC) under Grant no.61775113 and no.62175121 and the National Key Research and Development Program of China under Grant no. 2017YFF0206104. X.Z. acknowledges the support by the China Scholarship Council (CSC No. 201906210262). J.V acknowledges support under DARPA contract HR001118C0015. We thank Ms. Dongmei Hong from Beijing Qifeng Landa Optical Technology Development Co., Ltd. for the sample fabrication.

**References**

1. A. Marandi, Z. Wang, K. Takata, R. L. Byer, and Y. Yamamoto, "Network of time-multiplexed optical parametric oscillators as a coherent Ising machine," Nat Photonics **8**, 937–942 (2014).

2. T. Zhou, X. Lin, J. Wu, Y. Chen, H. Xie, Y. Li, J. Fan, H. Wu, L. Fang, and Q. Dai, "Large-scale neuromorphic optoelectronic computing with a reconfigurable diffractive processing unit," Nat Photonics 1–7 (2021).

3. J. Spall, X. Guo, T. D. Barrett, and A. I. Lvovsky, "Fully reconfigurable coherent optical vector–matrix multiplication," Opt Lett **45**, 5752 (2020).

4. P. Ambs, "Optical Computing: A 60-Year Adventure," Adv Opt Technologies **2010**, 1–15 (2010).

5. S. Abdollahramezani, O. Hemmatyar, and A. Adibi, "Meta-optics for spatial optical analog computing," Nanophotonics-berlin **9**, 4075–4095 (2020).

6. A. Silva, F. Monticone, G. Castaldi, V. Galdi, A. Alù, and N. Engheta, "Performing Mathematical Operations with Metamaterials," Science **343**, 160–163 (2014).

7. F. Zangeneh-Nejad, D. L. Sounas, A. Alù, and R. Fleury, "Analogue computing with metamaterials," Nat Rev Mater 1–19 (2020).

8. Y. Zhou, H. Zheng, I. I. Kravchenko, and J. Valentine, "Flat optics for image differentiation," Nat Photonics **14**, 316–323 (2020).

9. A. Cordaro, H. Kwon, D. Sounas, A. F. Koenderink, A. Alù, and A. Polman, "High-Index Dielectric Metasurfaces Performing Mathematical Operations," Nano Lett **19**, 8418–8423 (2019).

10. A. A. Dima, J. T. Elliott, J. J. Filliben, M. Halter, A. Peskin, J. Bernal, M. Kociolek, M. C. Brady, H. C. Tang, and A. L. Plant, "Comparison of segmentation algorithms for fluorescence microscopy images of cells: Comparison of Segmentation Algorithms," Cytom Part A **79A**, 545–559 (2011).

11. T. Falk, D. Mai, R. Bensch, Ö. Çiçek, A. Abdulkadir, Y. Marrakchi, A. Böhm, J. Deubner, Z. Jäckel, K. Seiwald, A. Dovzhenko, O. Tietz, C. D. Bosco, S. Walsh, D. Saltukoglu, T. L. Tay, M. Prinz, K. Palme, M. Simons, I. Diester, T. Brox, and O. Ronneberger, "U-Net: deep learning for cell counting, detection, and morphometry," Nat Methods **16**, 67–70 (2019).

12. N. Coudray, P. S. Ocampo, T. Sakellaropoulos, N. Narula, M. Snuderl, D. Fenyö, A. L. Moreira, N. Razavian, and A. Tsirigos, "Classification and mutation prediction from non–small cell lung cancer histopathology images using deep learning," Nat Med **24**, 1559–1567 (2018).

13. Y. LeCun, Y. Bengio, and G. Hinton, "Deep learning," Nature **521**, 436–444 (2015).

14. M. Le and S. Kayal, "Revisiting Edge Detection in Convolutional Neural Networks," Arxiv (2020).

15. W. T. Rhodes, "Incoherent Spatial Filtering," Opt Eng **19**, 193323-193323- (1980).

16. H. Wang, C. Guo, Z. Zhao, and S. Fan, "Compact Incoherent Image Differentiation with Nanophotonic Structures," Acs Photonics **7**, 338–343 (2020).

17. J. Jiang and J. A. Fan, "Global Optimization of Dielectric Metasurfaces Using a Physics-Driven Neural Network," Nano Lett **19**, 5366–5372 (2019).

18. P. Vogiatzis, S. Chen, X. Wang, T. Li, and L. Wang, "Topology optimization of multi-material negative Poisson's ratio metamaterials using a reconciled level set method," Comput Aided Design **83**, 15–32 (2017).

19. G. W Joseph, "Introduction to Fourier Optics, Roberts & Co," (n.d.).

20. V. Anand, J. Rosen, S. H. Ng, T. Katkus, D. P. Linklater, E. P. Ivanova, and S. Juodkazis, "Edge and Contrast Enhancement Using Spatially Incoherent Correlation Holography Techniques," Photonics **8**, 224 (2021).

21. T. B and S. S, "Bipolar Incoherent Image Processing for Edge Detection of Medical Images," International Journal of Recent Trends in Engineering **2**, 229 (2009).

22. T. Xu, J. He, H. Ren, Z. Zhao, G. Ma, Q. Gong, S. Yang, L. Dong, and F. Ma, "Edge contrast enhancement of Fresnel incoherent correlation holography (FINCH) microscopy by spatial light modulator aided spiral phase modulation," Opt Express **25**, 29207 (2017).

23. R. I. Rabady and A. Ababneh, "Global optimal design of optical multilayer thin-film filters using particle swarm optimization," Optik - Int J Light Electron Opt **125**, 548–553 (2014).

24. A. V. Tikhonravov, "Needle optimization technique: the history and the future," P Soc Photo-opt Ins 2–7 (1997).

25. "Modern design tools and a new paradigm in optical coating design," (n.d.).

26. Y. Shi, W. Li, A. Raman, and S. Fan, "Optimization of Multilayer Optical Films with a Memetic Algorithm and Mixed Integer Programming," Acs Photonics **5**, 684–691 (2017).

27. J. Jiang and J. A. Fan, "Multiobjective and categorical global optimization of photonic structures based on ResNet generative neural networks," Nanophotonics-berlin **10**, 361–369 (2020).

28. Y. Zhou, J. Zhan, R. Chen, W. Chen, Y. Wang, Y. Shao, and Y. Ma, "Analogue Optical Spatiotemporal Differentiator," Adv Opt Mater **9**, 2002088 (2021).

29. X. Qian, "Topology optimization in B-spline space," Comput Method Appl M **265**, 15–35 (2013).

30. S. J. Osher and F. Santosa, "Level Set Methods for Optimization Problems Involving Geometry and Constraints I. Frequencies of a Two-Density Inhomogeneous Drum," J Comput Phys **171**, 272–288 (2001).

31. T. Stangner, H. Zhang, T. Dahlberg, K. Wiklund, and M. Andersson, "Step-by-step guide to reduce spatial coherence of laser light using a rotating ground glass diffuser," Appl Optics **56**, 5427 (2017).

32. E. C. Kintner, "Investigations Relating to Optical Imaging in Partially Coherent Light," (1975).

33. P. Karimi, A. Khavasi, and S. S. M. Khaleghi, "Fundamental limit for gain and resolution in analog optical edge detection," Opt Express **28**, 898 (2020).

34. E. Moen, D. Bannon, T. Kudo, W. Graf, M. Covert, and D. V. Valen, "Deep learning for cellular image analysis," Nat Methods **16**, 1233–1246 (2019).

35. D. P. Hoffman, I. Slavitt, and C. A. Fitzpatrick, "The promise and peril of deep learning in microscopy," Nat Methods **18**, 131–132 (2021).

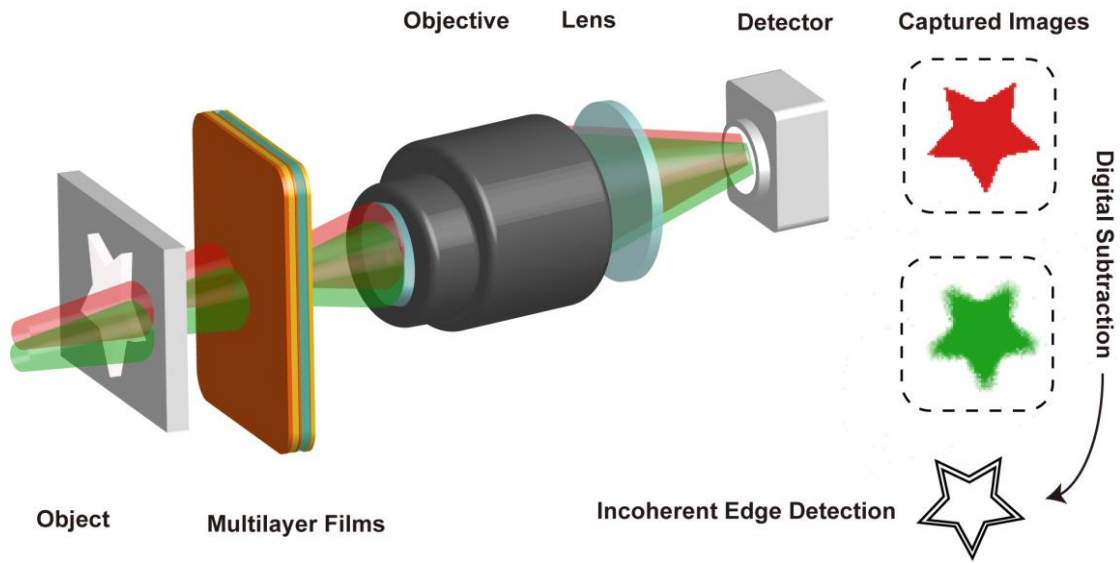

**Fig 1. Incoherent optoelectronic differentiation.** Schematic of the incoherent optoelectronic differentiation system with the multilayer films. The optical transfer functions of the multilayer films under both TE and TM polarizations at the wavelengths of 532 nm and 633 nm are deliberately engineered, so that the incoherent images at these two wavelengths appear slightly different with regard to the spatial frequency. The incoherent edge detection image is then acquired after the normalization and digital subtraction of the two incoherent images.

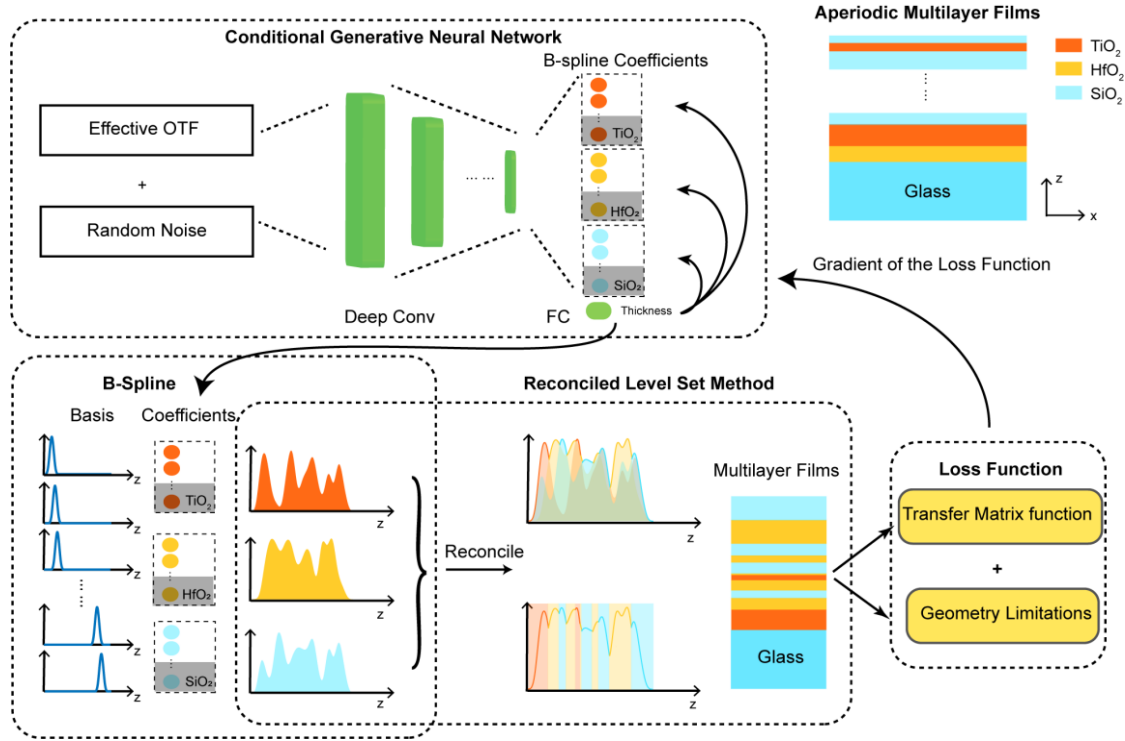

**Fig 2. Global optimization procedure of the multilayer films.** The conditional generative neural network maps the ideal effective OTF and a uniformly distributed noise vector to the B-spline coefficients for each material and a thickness factor. The coefficients are truncated according to the thickness factor and then used to build the level set function for each material. According to the reconciled level set method, each position is occupied by the material with the largest value of level set function at this certain position. TMM is then used to calculate the ATF of the established multilayer films at the wavelengths of 532 nm and 633 nm, based on which the effective OTF is obtained. The loss function is evaluated considering both the effective OTF and the geometric constraints of the multilayer films, and the gradients of the loss function with regard to the outputs of the neural network are then used to update the neuron weights in the neural network.

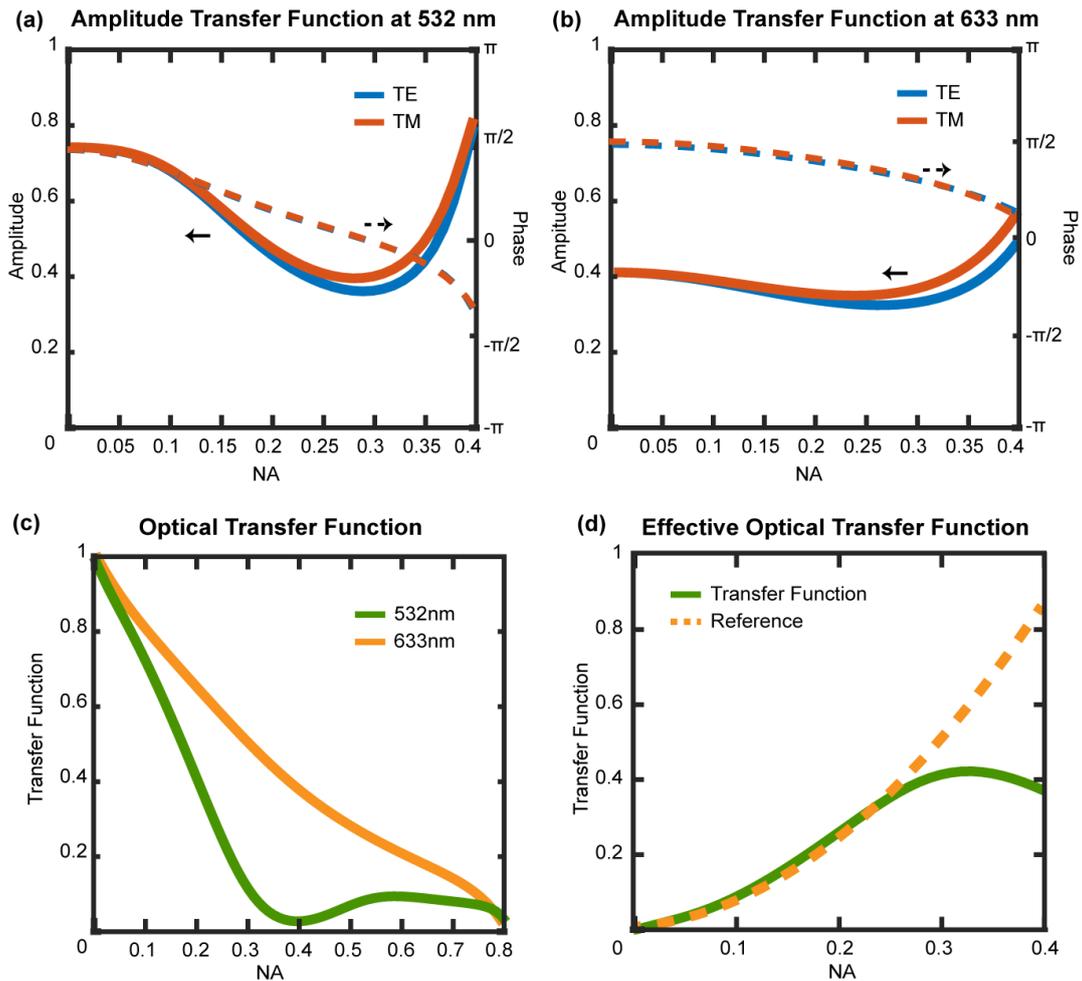

**Fig 3. Optical properties of the optimized multilayer films.** (a) The ATFs of the optimized multilayer films at 532 nm under both TE and TM polarizations. (b) The same as (a) but at the wavelength of 633 nm. (c) The OTFs of the optimized multilayer films at 532 nm and 633 nm. (d) The effective OTF of the optimized thin films, which is the subtraction of the two curves in (c). The quadratic curve is provided as a reference.

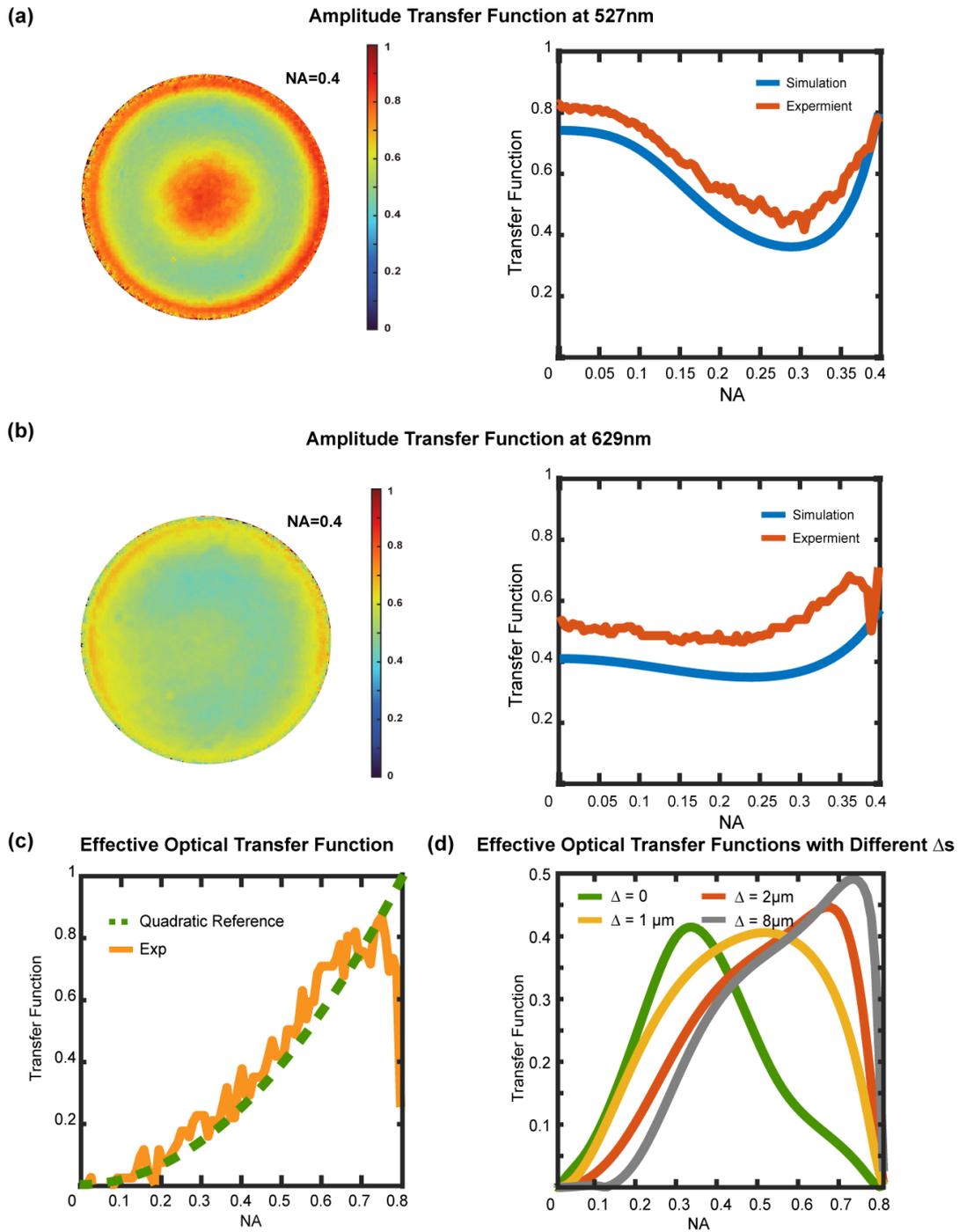

**Fig 4. Optical characterization of the fabricated multilayer films.** (a) Measured ATF at 527 nm and the comparison between the simulation and experimental results. (b) The same as (a) but at the wavelength of 629 nm. (c) Measured effective OTF together with the quadratic reference curve. (d) Simulated effective OTFs with different coherence. The results here assume an identical $\Delta$ at different wavelengths.

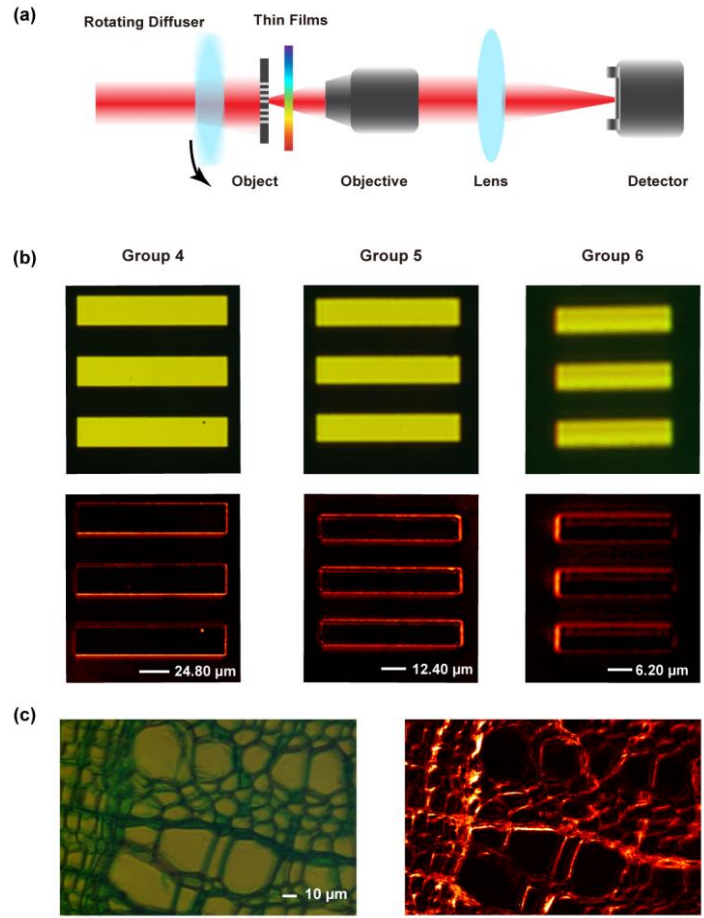

**Fig 5. Incoherent differentiation imaging using the fabricated multilayer films.** (a) Schematic of the experimental setup for incoherent optoelectronic differentiation imaging. (b) The captured RGB images of element 3 in Group 4 to Group 6 of 1951 USAF resolution test chart and the corresponding spatial differentiation results after electric subtraction. (c) The captured RGB image of the cross section of a woody dicot stem and its spatial differentiation result.